\documentstyle[aps,epsf,prb%,preprint,draft,tighten
  ]{revtex}\input {amssym.def}
\begin{document}
\title{ Dynamics of some Constrained Lattices}
\author{M.E. Simon, and C. M. Varma}
\address{Bell Laboratories, Lucent Technologies \\
Murray Hill, NJ 07974}
\maketitle

\begin{abstract}
We consider the dynamics of lattices which have constrained 
constitutive units flexible in only their mutual orientations.
A continuum description is derived through which it is shown that
the models have zero shear velocity, free-particle like internal rotational modes
and volume decreasing linearly with temperature. The relevance of models to
a range of problems is pointed out.

\end{abstract}


\section*{Introduction}
The dynamical properties of solids with strong anharmonicity are often 
usefully studied by specifying the constraints on the atoms rather than the forces
 between them. This approach has yielded interesting results in phenomena 
ranging from the properties of interfaces \cite{neumann}
to criteria for glass-formation \cite{phillips,thorpe} to rigidity percolation 
\cite{thorpe2}
in solids with "floppy Modes".

 We consider here the problem of lattices in which 
some  (strongly anharmonic) bonds can be deformed at a
much lower energy than the others. As noted below the problem is of wide interest 
but our attention to it is drawn by
the $contraction$ in volume of a class of solids  $linearly$ as temperature
is increased, in a range of temperature which in some cases extends from $%
\sim $ 10K to the highest measured temperature \cite{mary,ram1}. Conventional perturbative 
calculations 
with anharmonic potentials are not very helpful in this case. These solids have a
complicated lattice structure but always contain polyhedra which share some
or all corners with the neighbors. Any distortion of the polyhedra is
expensive, the vibrations changing the angles between the polyhedra however
have low energies $\simeq $ $\omega _{0}.$ 
 A simple model can be constructed
to show the phenomena of thermal contraction in such cases. A
two-dimensional version is shown in fig. 1a, which consist of a solid with
three squares per unit cell, two of which are connected only at two corners
(these can be replaced by sticks of fixed length) while the third is
connected at all four corners \cite{alternate}. All bond-lengths as well as the internal
right-angles of the squares are constrained to be fixed. The only independent 
variable
is the angle $\theta _{i}$ between the squares.

The area occupied by the solid is the sum of the area occupied by the
solid squares and the area between them. The latter are bounded by perimeters
whose length are fixed. Under this condition the area is least when the
perimeter has the highest symmetry, i. e. in the configuration shown in fig.
1a. Any variation in $\left\{ \theta _{i}\right\} $ must then necessarily
lead to a reduction in volume. For the unit cell shown, the area decreases
as $ \theta _{i}^{2}$. We expect $\langle  \theta ^{2}\rangle
\sim T$ for $T>\omega _{0}$, giving the observed behavior \cite{conventional,footnote3}.

The above is the correct dimensional argument for the thermal contraction,
but it should be clear that $\theta $ 's in different cells are
coupled, and coupled non-locally to the strain. The coupling of translational 
and rotational degrees of freedom leads in the continuum limit to a
 field-theoretical problem with some remarkable features. Besides the thermal 
contraction, we have found that the shear modes have zero stiffness in such solids
and that they have rotational modes with dispersion
characterestic of free particles: $\omega \propto k^2$.  These features are
also present in three-dimensional versions of the model and are expected 
also to hold in more general models than studied here. 

A convenient set of variables for the problem are the position of the four
corners of each square $i$, $\left\{ {\bf \Psi }_{i,n}={\bf R}_{i}+{\bf r}%
_{i,n}\right\} $ , with ${\bf R}_{i}$ the coordinate of the center of the
square $i$ and $n=\pm 1,\pm 2$ label each corner (see Fig. 1b). Fixing the lengths and the
internal right-angles of the squares, the relatives positions ${\bf r}_{n,i}$
can be expressed as functions of the angle $\theta _{i}$ alone, ${\bf r}_{i,n}=b(cos{(\theta _{i}+(n-1)\pi /2)}\hat{x}+sin{(\theta _{i}+(n-1)\pi /2)} \hat{y})$, where $b$ is a
fixed length. The original 8 degrees of freedom per unit cell are then
reduced to 3 $\left\{ {\bf u}_{i},\theta _{i}\right\} $, where ${\bf u}_{i}=%
{\bf R}_{i}-{\bf R}_{i}^{0} \equiv (u_{ix}\hat{x}+u_{iy}\hat{y})$ 
is the displacement of the center of mass of
the squares around the equilibrium position at $T=0$. Two more constraints per
unit cell come from  fixing  the lengths of each connecting stick (or
square);

\begin{equation}
\left| {\bf \Psi }_{i,1}-{\bf \Psi }_{i+x,-1}\right| =\left| {\bf \Psi }%
_{i,2}-{\bf \Psi }_{i+y,-2}\right| =2b  \label{cons}
\end{equation}
the labels $i\pm x$, and $i\pm y$ represent the neighbors of the cell $i$.

In any distortion about the symmetric position, the two-diagonals in each
square change angle with respect to the connecting sticks . The potential
energy may be expressed in terms of the four such angles per unit cell, 
\begin{equation}
V=K\sum_{i,n=\pm 1,\pm 2}(1-\cos \mu_{i,n}),  \label{v1}
\end{equation}
$\mu_{i,n}$ may be written in terms of the vectors ${\bf \Psi }_{i,n}$.
For example
\begin{equation}
\cos \mu _{i,1}=\frac{
({\bf \Psi }_{i,1}-{\bf \Psi }_{i,-1}) \cdot ({\bf \Psi }_{i+x,-1}-{\bf \Psi }%
_{i,1})}{\left| {\bf \Psi }_{i,1}-{\bf \Psi }_{i,-1}\right| {\bf \ }\left| 
{\bf \Psi }_{i+x,-1}-{\bf \Psi }_{i,1}\right| }
\end{equation}

In the continuum limit $a\partial _{\alpha}{\bf r\ll 1,}$ $a\partial _{\alpha}
{\bf %
u\ll 1}$ ($a=4b,\alpha=(x,y))$, the potential energy simplifies to 
\begin{equation}
V=4K\int \left[ 2-\frac{1}{b}%
\left( {\bf r}_1 \cdot (2 \hat{x}+\partial _{x}{\bf u})
+{\bf r}_2 \cdot (2 \hat{y}+\partial _{y}{\bf u}))
+\frac{1}{2}({\bf r}_1%
\cdot \partial_x^2 {\bf r}_1+{\bf r}_2\cdot \partial_y^2 {\bf r}_2)\right) \right] dS  \label{V}
\end{equation}

Note that terms quadratic in $\partial_{\alpha}\bf{u}$ do not occur in Eq. ( \ref{V} ). This Equation can be written as: 
\begin{equation}
V=4K\int \left[ 2-\cos \theta \ (2+\nabla {\bf u)}-\sin \theta \ (\nabla
\times {\bf u})-2b^{2}(\nabla \theta {\bf )}^{2}\right] dS. \label{Vbis}
\end{equation}
The rotational invariance of Eqs. (\ref{V}) and (\ref{Vbis}) will be evident
 only after the constraints of Eq. (\ref{cons}) are implemented.

The kinetic energy is $T=\frac{1}{2}m\sum_{i,n}(\stackrel{\bullet }{\bf \Psi }%
_{i,n})^{2}$ which in the continuum limit becomes 
\begin{equation}
T=\frac{1}{2}\int (M(\stackrel{\bullet }{\bf u})^{2}+4mb^{2}(\stackrel{%
\bullet }{\theta })^{2})dS  \label{T}
\end{equation}

To obtain the continuum version of the constraints of Eq.(1), consider all 
the four sticks connecting any given square to its neighbors. The four 
equations can be written as
 
\begin{eqnarray}
\left| {\bf A}_{\alpha }\right| ^{2}+\left| {\bf B}_{\alpha }\right| ^{2}
\pm 2{\bf A}_{\alpha } \cdot {\bf B}_{\alpha }
&=&4b^{2}  \nonumber \\
or
\left| {\bf A}_{\alpha }\right| ^{2}+\left| {\bf B}_{\alpha }\right| ^{2}
&=&4b^{2}  \nonumber \\
{\bf A}_{\alpha }\cdot {\bf B}_{\alpha } &=&0  \label{ab0}
\end{eqnarray}
where, 
\begin{eqnarray}
{\bf A}_{\alpha } &=&a(\hat{\alpha }+\partial _{\alpha }{\bf u)-}2{\bf r-}%
\frac{a^{2}}{2}\partial _{\alpha }^{2}{\bf r}  \nonumber \\
{\bf B}_{\alpha } &=&\frac{a^{2}}{2}\partial _{\alpha }^{2}{\bf u-}a\partial
_{\alpha }{\bf r,}  \label{AB}
\end{eqnarray}

These four equations determine the four components of the strain tensor
$\partial_{\alpha}\bf{u}$.
We have found that these constraints are satisfied to $O(\theta ^{2})$ by
two sets of relations, as may be verified by substitution

Set I: 
\begin{eqnarray}
a(\hat{x}+\partial _{x}{\bf u)} &=&4{\bf r}_1+\frac{3}{4}a^{2}\partial _{x}^{2}{\bf r}_1 ,\nonumber \\
a(\hat{y}+\partial _{y}{\bf u)} &=&4{\bf r}_2+\frac{3}{4}a^{2}\partial
_{y}^{2}{\bf r}_2.  \label{set1}
\end{eqnarray}
Set II: 
\begin{eqnarray}
a(\hat{x}+\partial _{x}{\bf u)} &=&2({\bf r}_1+{\bf r}_1^{\prime})+\hat{y} f_{x} +a^{2}(\frac{%
1}{4}\partial _{x}^{2}{\bf r}_1+\frac{3}{4}\partial _{x}^{2}{\bf r}_1^{\prime}) 
\nonumber \\
a(\hat{y}+\partial _{y}{\bf u}) &=& 2({\bf r}_2+{\bf r}_2^{\prime})-
\hat{x} f_{y} +a^{2}(%
\frac{1}{4}\partial _{y}^{2}{\bf r}_2+\frac{3}{4}\partial _{y}^{2}{\bf r}_2^{\prime})  \label{set2}
\end{eqnarray}

where ${\bf r}_n^{\prime}=b(cos{(\theta+(n-1)\pi /2)}\hat{x}-sin{(\theta +(n-1)\pi /2)} \hat{y}) $and $f_{x}(\theta )$ and $f_{y}(\theta )$ are arbitrary real functions of
order $\theta ^{2}$ or higher. 
These relations relates the deformation of the {\it unit cell} with $\theta $%
. For Set I the first term on the right side rigidily rotates each unit cell
by $\theta $, a uniform $\theta $ gives a rigid rotation of the entire
solid. The second term on the right hand allows a smooth spatial variation
of $\theta $ (see Fig. 2). The deformation for the set II are shown in Fig.
1b. In these optical-like modes the connecting stick rotates by $-\theta $
and the cell contracts $\sim \theta ^{2}$ in each direction.

Now we determine the equation of motion for each set of solutions.
Eqs. (\ref{set1}) can be combined in the form 
\begin{eqnarray}
{\bf \nabla \cdot u} &=&2(\cos \theta -1)-3b^{2}\nabla.\nabla(\cos \theta)
\equiv J_{0}(\theta ),  \nonumber \\
{\bf \nabla \times u} &=&2\sin \theta +3b^{2}\nabla.\nabla(\sin\theta)
\equiv J_{1}(\theta ).
\label{J1}
\end{eqnarray}
${\bf J}_{0}(\theta )$ and ${\bf J}_{1}(\theta )$ are thus the sources
determining the irrotational and solenoidal parts of ${\bf u}$. The other
 two equations obtained from Eqs.(\ref{set1}) are that the shear strain
$(\partial_{x}u_y + \partial_{y}u_x)=3b^2(\partial^2_x-\partial^2_y)(sin\theta)$,
and $(\partial_xu_x-\partial_yu_y)= 3b^2(\partial^2_x-\partial^2_y)(cos\theta)$.
For results at long wavelengths these can be equated to zero.
 { \it The model thus has zero shear-velocity}.

With Eqs. (\ref{set1}), it is straight forward to show that 
\begin{equation}
V=-4Kb^{2}\int \left| \nabla \theta \right| ^{2}dS.  \label{V1}
\end{equation}

The absence of terms proportional to $\theta ^{2}$ in (\ref{V1}) has
important consequences. Their absence may be traced to the local invariance
satisfied by (\ref{J1}):
\begin{equation}
\left| \frac{{\bf \nabla \times u}}{2}\right| ^{2}+\left| 1+\frac{{\bf %
\nabla \cdot u}}{2}\right| ^{2}+3b^{2}\left| \nabla \theta \right| ^{2}=1
\label{inv}
\end{equation}
Corrections to Eq. (\ref{inv}) are of $O(b^4(\nabla\theta)^4)$.

Consider next the equation of motion for $\theta(\bf{r},t)$. To do this
${\bf u}$ in the kinetic energy $T$ must be expressed in terms of the sources 
${\bf J}%
_{0}(\theta )$, ${\bf J}_{1}(\theta )$. Noting that $\left| {\bf J}%
_{0}(\theta )\right| ^{2}+\left| {\bf J}_{1}(\theta )\right| ^{2}=4\theta
^{2}+12b^{2}\theta \nabla ^{2}\theta +O(\theta ^{3}),$ the equation of motion
for $\theta$
to linear order associated with set I is found to satisfy the energy-wave
vector relation
\begin{equation}
(m-6M)\omega ^{2}-4Kq^{2}+4M\frac{\omega ^{2}}{q^{2}}=0 
\label{disp}
\end{equation}
The last term comes from $(\stackrel{\bullet }{\bf u})^{2}$. Eq. (\ref{disp}) 
yields the interesting result that at long-wave-length, free-particle like
modes exist for which $\omega =\frac{K}{M}q^{2}$. The physical reason for
the existence of such rotational mode lies in the local invariance (\ref{inv}) 
and the
fact that rotation of the entire solid ($q=0$) involves transporting
macroscopic mass. At long-wavelengths
the effective mass is $\sim M/q^{2}.$
The characterestic frequency below which the dispersion of the modes
 is proportional to $q^2$,
determined by looking at corrections of $O(q^4))$ in Eq.(\ref{disp}), is $O(K/M)^{1/2}$.

For Set II the potential energy is given by 
\[
V=4K\int [1-\cos 2\theta \ (1{\bf +}b^{2}\left| \nabla \theta \right|
^{2})+\sin \theta \ (f_{x}-f_{y})]\ dS
\]

and the equations for the sources of the fields for $\nabla.\bf{u}$
and $\nabla \times \bf{u}$ are, respectively 
\begin{eqnarray}
J_{0}(\theta ) &=&2\cos \theta -2-3b^{2}(\cos \theta \ (\nabla \theta
)^{2}+\sin \theta \ \nabla ^{2}\theta ),  \nonumber \\
J_{1}(\theta ) &=&f_{x}-f_{y}+b^{2}(\cos \theta \ \nabla ^{2}\theta -\sin
\theta \ (\nabla \theta )^{2}).  \label{J2}
\end{eqnarray}

The contribution of the source-fields to the kinetic energy 
is now $\sim \theta ^{4}$ and
does not change the equation of motion for $\theta(\bf{r},t)$
which give the expected
 optical-like
dispersion relation 
\begin{equation}
\omega ^{2}=\frac{8K}{M}(1-\frac{b^{2}q^{2}}{2}). 
\end{equation}
To linear order $f_\alpha$ do not enter in the dispersion.

Since the problem has only one degree of freedom per 
unit cell, the free-particle like and the optical-like modes
are not linearly independent. This
may be seen through Fig. 2. The free-particle
like modes can be locally thought of as  composed of a rotation by $\theta
+\delta \theta $ and a $q=\pi /a$ optical mode of amplitude $\delta \theta
/2.$ The sharing of the spectral weight between collective modes
in non-linear problem is known in other contexts.

Now we come to the question of thermal compression with which we started. 
The equilibrium volume of the lattice at a temperature T may be determined
from the thermal expectation value 
\begin{equation}
\langle \nabla.\bf{u} \rangle = \langle J_0(\theta) \rangle.
\end{equation}
From Eqs. (\ref{J1}) and (\ref{J2}), we note that both sets of modes
 contribute to the
thermal contraction at all temperatures which to the order of the theory
here is proportional to the expectation value $-\langle
\theta^2(T) \rangle$. 
For T larger than $O(\omega_0) = 
 O(K/M)^{1/2}$ the thermal contraction may then
 be determined by the equipartition theorem to be proportional to T. The thermal contraction is entropic akin to that of rubber.

The problem can be extended to 3 dimensions replacing the squares by
octahedra and adding an extra stick per cell which connects the corners of
neighboring octahedra in the $z$ direction. The model has a cubic symmetry.
For a fixed size of the octahedra  the 6 degrees of freedom per unit cell (3
translation and 3 rotations) are reduced to 3 by fixing the length of the
sticks (analogues of Eq.(1)). the 
coordinates of the 6 vertices of each octahedron can again be expressed as 
$\bf \Psi_{i,n}={\bf R}_{i}+{\bf r}_{i,n}$, with ${\bf r}_{i,1}=-{\bf r}%
_{i,-1}=(b,0,0)\widehat{C}_{\phi \theta \psi }$, ${\bf r}_{i,2}=-{\bf r}%
_{i,-2}=(0,b,0)\widehat{C}_{\phi \theta \psi }$ and ${\bf r}_{i,3}^{\dagger
}=-{\bf r}_{i,-3}^{\dagger }=(0,0,b)\widehat{C}_{\phi \theta \psi }$ where $%
\widehat{C}_{\phi \theta \psi }=\widehat{C}_{\psi }^{\widehat{z}}$ $\widehat{%
C}_{\theta }^{\widehat{x}}\widehat{C}_{\phi }^{\widehat{z}}$ represent  the
three rotations by the Euler angles $\phi $, $\theta $ and $\psi $ \cite
{gold}.

The 3 constraints in the continuum limit can  be expressed in the same form
as Eq.  (\ref{ab0}), where now 
\begin{eqnarray*}
{\bf A}_{\alpha } &=&a(\widehat{e}_{\alpha }+\partial _{\alpha }{\bf u)}-2%
{\bf r}_{n_{\alpha }}{\bf -}\frac{a^{2}}{2}\partial _{j}^{2}{\bf r}%
_{n_{\alpha }} \\
{\bf B}_{\alpha } &=&\frac{a^{2}}{2}\partial _{\alpha }^{2}{\bf u-}a\partial
_{\alpha }{\bf r}_{n_{\alpha }}.
\end{eqnarray*}
${\bf A}_{\alpha }$ and ${\bf B}_{\alpha }$ are three component vectors, 
$\alpha = (x,y,z)$, $\widehat{e}_{\alpha }$ is the unit
 vector in the $\alpha$ direction and $n_{\alpha}=(1,2,3)$, indicate the 
corresponding corners of the octahedra. Once again the shear velocities
are zero.
 Also we again get two sets of solutions.

Set I:

\begin{equation}
a(\widehat{e}_{\alpha }+\partial _{\alpha }{\bf u)}=4{\bf r}_{n_{\alpha }}%
{\bf +}\frac{3}{4}a^{2}\partial _{\alpha }^{2}{\bf r}_{n_{\alpha }}
\label{set13d}
\end{equation}

Note that ${\bf r}_{n_{\alpha }}$ are functions of the Euler angles, then
again eqs. (\ref{set13d}) give the deformation of each unit
cell as a function of the Euler angles. 

The potential energy can be found analogous to the 2D case (Eq. \ref{V1})
and results

\begin{equation}
V_{3D}=-b^{2}8K\int \left[ Tr\left( \widehat{C}_{\phi \theta \psi }^{-1}\cdot \nabla ^{2}\widehat{C}_{\phi \theta \psi
}\right) \right] dV
\end{equation}

with $\left( \nabla ^{2}\widehat{C}_{\phi \theta \psi }\right)
_{ij}=\partial _{i}^{2}(\widehat{C}_{\phi \theta \psi })_{i,j}.$ The absence
of a mass term is again due to an equations analogous to Eqs. (\ref{inv}). We
again obtain 3 modes with $\omega \sim q^{2}$ for reasons similar to the
2-dimensional case.

For Set II 
\begin{equation}
a(\widehat{e}_{\alpha }+\partial _{\alpha }{\bf u)}=2({\bf r}_{k_{j}}+{\bf r}%
_{k_{j}}^{\prime})+a^{2}(\frac{1}{4}\partial _{x}^{2}{\bf r}%
_{n_{\alpha }}+\frac{3}{4}\partial _{x}^{2}{\bf r}_{n_{\alpha }}^{\prime}),
\end{equation}

where is an arbitrary vector satisfying $|{\bf r}_{n_{\alpha }}^{\prime}|=2b
$. From here it can be  derived the three optical modes. 

We have also considered the simple  the model in which the sticks connecting the rigid  polyhedra (squares in 2-dimension) have finite elasticity, i.e. the constraints of Eqs. (\ref{cons}) are not imposed. The model then has of course the usual
longitudinal and transverse modes at long wavelengths. The results obtained above cannot be obtained in any limit of the parameteres of such harmonic model as a pertubative expansion about them. The constraints lead to qualitativaly new class of properties. These properties are preserved if wegeneralize the sticks in Fig. (2) to a polymer, i.e. with a chain of fixed length sticks deformable in the mutual orientation at the connecting points.
  
The model investigated is especially interesting  in some unfamiliar contexts. It has direct application to sintered solids \cite{sintered},
 where rigid 
particles are connected to each other through bonds with small energies for angular
variations although the model needs to be generalised 
to take into account the 
random placements of the bonds. The predictions for zero or very small shear velocity 
and of long wavelength rotational modes is expected to hold generally and should be tested in that case.

Some models for granular solids\cite{edwards} are related to the model 
studied here again with
generalisation to the random placing of the bonds. The prediction of zero shear
 velocity
obtained here is particularly interesting in connection with granular materials
because they flow like a liquid.\cite{nagel} Of-course to examine flow the model must 
also be generalised to allow for "bond-breaking". The model studied here is also a particular example of general models discussed by S. Alexander \cite{alex} with potential applications to unusual excitation modes in glasses and foams.

\section*{acknowledgments}

We thank Glenn Kowach and Art Ramirez for introducing 
us to the problem of thermal contaction and Sharad Ramnathan, Steve Simon and 
Alexei Tkachenko for very useful discussions.
M.E.S is a fellow of the Consejo Nacional de Investigaciones Cient\'{\i
}ficas y T\'{e}cnicas (CONICET), Argentina.

\*{FIGURE CAPTIONS}

{\bf Figure 1} The unit cell of the model of rigid squares hinged at corners. (a) At the minimun of potential, (b) with a given distortion exhibiting compression in the optical mode.

{\bf Figure 2}  exhibits the distortions in the rotational mode.


\begin{references}

\bibitem{neumann} see for example J. Von Neumann in {\it Metal Interfaces},
  Cherrin, ed., ASM Cleveland OH (1952), p108;
D. Wearie and S. Mc Murray, { \it Solid State Physics}, Ehrenreich and Spaepen, eds., Vol 50 (1996).
\bibitem{phillips} J.C. Phillips, J.Non-Cryst. Solids {\bf 34},153 (1979).
\bibitem{thorpe} M.F. Thorpe, J. Non-Cryst. Solids {\bf 57},355 (1983).
\bibitem{thorpe2} M.F. Thorpe, Phys. Rev. Lett. {\bf 75},4051 (1995).
\bibitem{mary}  T. A. Mary, J.S.O. Evans, T. Vogt, and A.W. Sleight, 
Science 
{\bf 272}, 90 (1996).
\bibitem{ram1}  A. P. Ramirez and G. R. Kowach,
 Phys. Rev. Lett., {\bf {80}}%
, 4903, (1998); G Ernst, C. Broholm, G. Kowach, and A. P. Ramirez, Nature (London) {\bf 396}, 147 (1998).
\bibitem{alternate} A model in which all (rigid) squares are
 four-fold connected
has only one degree of freedom for the entire solid.
\bibitem{conventional} The conventional wisdom that a harmonic 
solid has no thermal
expansion or contraction is circumvented in our model by the 
constraints
which in effect enforce the effects of anharmonicity. In a theory obeying 
the constraints for some degrees of freedom a harmonic theory for the remaining degrees 
of freedom gives thermal-expansion or contraction.
\bibitem{footnote3} Models in which rigid rotation of polyhedra is invoked to
obtain thermal contraction have been considered before \cite{pry}. The mathematical 
implementation of the idea is done however in models of the type 
mentioned in footnote (6) by splitting the corner atom into two, which misses the new features of the problem.
\bibitem{pry}  A. K. A. Pryde, et. al., Phys. cond. matt. 8 ,10973 (1996); and references therein.
\bibitem{gold}  Golstein, {\it Classical Mechanics}, Addison-Wesley Publishing Company, Inc, p. 107 (1959).
\bibitem{edwards} S.F.Edwards and D.V.Grinev, Phys. Rev. Lett.
{\bf 82},5397 (1999).
\bibitem{nagel} H.M Jaeger, S. R. Nagel and R.P. Behringer, Rev. Mod. Phys. {\bf 68}, 149 (1996).
\bibitem{sintered} see for example {\it The Physics of  Powder Metallurgy}, Walter E Kignston, ed., Mc Graw Hill (1951).  
\bibitem{alex} S Alexander, Phys. Reports, {\bf 296}, 65 (1998).

\end{references}
\end{document}